\documentclass[aps,pra,twocolumn]{revtex4}

\usepackage{amssymb}
\usepackage{graphicx}
\usepackage{amsmath}
\usepackage{dcolumn}
\usepackage{bm}

\newcommand{\beq}{\begin{equation}}
\newcommand{\eeq}{\end{equation}}
\newcommand{\beqa}{\begin{eqnarray}}
\newcommand{\eeqa}{\end{eqnarray}}

\newcommand{\la}{\langle}
\newcommand{\ra}{\rangle}
\newcommand{\bE}{{\bf E}}
\newcommand{\bx}{{\bf x}}
\newcommand{\by}{{\bf y}}

\newcommand{\bu}{{\bf u}}

\def\ka#1{{\kappa}_{#1}}

\def\ajp#1{{ Am.\ J.\ Phys.} {\bf #1}}

\def\jpb#1{{ J.\ Phys.\ B} {\bf#1}}

\def\lphys#1{{ Laser\ Phys.} {\bf#1}}
\def\nat#1{{ Nature} {\bf#1}}

\def\oc#1{{ Opt.\ Commun.} {\bf#1}}
\def\ol#1{{ Opt.\ Lett.} {\bf#1}}
\def\pla#1{{ Phys.\ Lett. A\/} {\bf#1}}

\def\pra#1{{ Phys.\ Rev. A\/} {\bf#1}}

\def\pre#1{{ Phys.\ Rev. E\/} {\bf#1}}
\def\prl#1{{ Phys.\ Rev.\ Lett.} {\bf#1}}

\begin{document}

\title{Entanglement and Classical Polarization States}

\author{Xiao-Feng Qian}
\email{xfqian@pas.rochester.edu}
\author{J.\ H.\ Eberly}
\affiliation{ Rochester Theory Center and the Department of Physics
\& Astronomy\\
University of Rochester, Rochester, New York 14627}

\begin{abstract}
We identify classical light fields as physical examples of
non-quantum entanglement.  A natural measure of degree of
polarization emerges from this identification, and we discuss its
systematic application to any optical field, whether beam-like or
not.
\end{abstract}

\maketitle

\noindent From Christian Huygens' explanation of the fascinating
birefringent property of the crystals called Iceland spar (i.e.,
calcite), through the much later work of Sir George Stokes
\cite{Stokes}, the formulation of polarization theory has
continuously evolved and is now well established in terms of field
correlation functions \cite{statoptics}. The concept of degree of
polarization is based on the coherence matrix or polarization matrix
constructed from these functions. But, even after centuries of
attention to this basic property of optical fields, fascinating new
issues concerning polarization have emerged in the past two decades.

The familiar measures of polarization come from the treatment of
light as a beam. This implies a given direction of propagation, and
thus a specific transverse plane. But development of highly
non-paraxial fields, use of very narrow-aperture imaging systems,
recognition of associated propagation questions, and probing of
fully three-dimensional fields as in hohlraums, all point to
necessary modifications of polarimetry beyond the traditional
picture \cite{Ash-Nicholls, Pohl-etal84, Alonso-etal}.

Questions of definition and of principle are also in need of
answers. For an electromagnetic field without a clear transverse
plane or even cylindrical symmetry the relevance of measures such as
the conventional degree of polarization must be reconsidered
\cite{Brosseau}, and alternative approaches have been proposed
\cite{Samson73, Carozzi00, Klimov-etal05, Luis, Friberg-etal,
Ellis-etal}. The same is true of experimentally important measures,
such as the Mueller matrices \cite{Kumar-Simon, Simon-etal10}. In
addition, pioneering measurements \cite{Masalov, Tsegaye-etal,
Klimov-etal10, Iskhakov-etal11} have highlighted two-photon and
multi-photonic views of polarization \cite{Klyshko92, Klyshko97},
and polarization measures for nonlinear classical optical waves have
been considered \cite{Picozzi}. Here we describe a basis for
comprehensive consideration of interrelated questions by addressing
the issue of dimensionality in a unified way.

Of course, dimensionality is trivially engaged in converting the
field from planar-transverse to non-planar, which only requires the
natural extension \beqa \label{E} {\bf E} = {\bf x}E_{x} + {\bf
y}E_{y} \Longrightarrow  {\bf x}E_{x} + {\bf y}E_{y} + {\bf z}E_{z},
\eeqa in order to take into account a third component of the field.
However, an entirely comprehensive treatment of polarization
questions should begin by first noting that two independent vector
spaces are employed in each realization of ${\bf E}$, and second
that the two spaces are entangled \cite{Qian-Eberly}. Entanglement
as a technical term means just that ${\bf E}$, in (\ref{E}), is a
tensor product of ``lab space" unit vectors such as $\bx$ and $\by$,
and functions $E_{x}$ and $E_{y}$ that are vectors in a statistical
``function space" of continuous normed functions (typically taken to
depend on space-time or space-frequency).

The light field uses at least one vector from each of these distinct
spaces, and in the general case it is not possible to convert
(\ref{E}) to the completely polarized form ${\bE} = {\bf u}F$, in
which the two spaces appear only in a factored direct product. Such
a (typically unlikely) form is called by the equivalent terms
``separable" or ``factorable" or ``non-entangled". That is,
determining degree of polarization is the same as determining degree
of factorization (separability) of the two spaces, i.e., absence of
entanglement. This fact can be exploited.

Both inner and outer products play a role in polarization theory.
For convenience we first consider a beam and write the fundamental
polarization outer product, \beqa \label{E><E} |{\bf E}\ra\la{\bf
E}|  = \Big(|{\bf x}\ra|E_x\ra + |{\bf y}\ra|E_y\ra \Big) \Big(
\la{\bf x}|\la E_x| + \la{\bf y}|\la E_y| \Big), \eeqa as the field
intensity times a normalized hermitian outer product ${\cal W}$,
i.e., $I\ {\cal W} = |{\bf E}\ra\la{\bf E}|$. Here the angle
brackets explicitly make the point mentioned above that the unit
vectors and the field components are members of different vector
spaces. This is a quantum-like notation that will be helpful, but no
quantum properties will be introduced.

With the intensity ${I} = \la E_x|E_x\ra + \la E_y|E_y\ra$ factored
out, we can write \beqa \label{E-sincos} |\bE\ra =
\sqrt{I}\Big(\cos\theta|\bx\ra|e_x\ra + \sin\theta|\by\ra|e_y\ra
\Big), \eeqa and this allows ${\cal W}$ to be written: \beqa
\label{W-e} {\cal W} &=& \Big(\cos\theta|\bx\ra|e_x\ra +
\sin\theta|\by\ra|e_y\ra \Big)
\Big(\la\bx|\la e_x| \cos\theta \nonumber \\
&& + \la\by|\la e_y|\sin\theta \Big), \eeqa where assignment of the
relative amplitudes via sine and cosine factors allows the
components $|e_x\ra$, $|e_y\ra$ to be unit-normalized: $\la
e_i|e_i\ra = 1$. We take account of the generally non-zero
correlation between the field's components by introducing the
magnitude and phase of the cross correlation as \beq
\label{alphaDef} \la e_x|e_y\ra \equiv \alpha = |\alpha|
e^{i\delta}. \eeq

Although rarely mentioned, each of the two separate vector spaces
has its own polarization matrix. These are reduced-state tensors,
i.e., traced over one space independent of the other. The ``normal"
polarization matrix is obtained by tracing over the function space.
We can denote it as ${\cal W}_{\rm lab} = Tr_{\rm fcn}({\cal W})$,
and calculate it by a diagonal sum over any complete set of
orthonormal vectors in the function space (but only in the function
space), a set that we can label $\{|\phi_m\ra \}$. For short we will
temporarily use $p$ and $q$ to stand for $x$ or $y$, and then obtain
\beqa \label{W_lab} I\,{\cal W}_{\rm lab} && = I\,Tr_{\rm fcn}({\cal
W}) =
\Sigma_m\la\phi_m|{\bf E}\ra\la {\bf E}|\phi_m\ra \nonumber \\
&&= \Sigma_{p,q}|{\bf p}\ra\la{\bf q}|\la E_q|E_p\ra, \ (p,q = x\
{\rm or}\ y), \eeqa which we recognize as a $2 \times 2$ tensor in
the lab space in the basis defined by $|\bx\ra$ and $|\by\ra$. When
the intensity is factored out we have a familiar matrix expression
with the required unit trace: \beqa \label{Wlab} {\cal W}_{\rm lab}
= Tr_{\rm fcn}({\cal W}) = \left[
\begin{array}{cc}
\cos^2\theta & \alpha\cos\theta\sin\theta \\
\alpha^*\sin\theta\cos\theta & \sin^2\theta
\end{array}
\right] \eeqa

The less familiar other polarization matrix is obtained by tracing
over the lab space. This is trivially done via the projections
$|\bx\ra\la\bx|$ and $|\by\ra\la\by|$, but the result will not be in
standard form because of the non-correlation of $E_x$ and $E_y$. We
can overcome this by rewriting $|\bE\ra$ as a sum of a pair of
statistically orthogonal components. If we choose $|e_x\ra$ as one
of them, we will denote its partner by $|\bar e_x\ra$, with $\la\bar
e_x|e_x\ra \equiv 0$. Then $|e_y\ra$ becomes a combination of both
components, $\alpha|e_x\ra + \beta|\bar e_x\ra$, so \beqa |\bE\ra
/\sqrt{I} = (\cos\theta|{\bx}\ra + \alpha\sin\theta|{\by}\ra)|e_x\ra
+ \beta\sin\theta|{\by}\ra|\bar e_x\ra, \eeqa where $\alpha$ is
defined in (\ref{alphaDef}), and $|\alpha|^2 + |\beta|^2 =1$. Then
the reduced polarization tensor for the function space ${\cal
W}_{\rm fcn} = Tr_{\rm lab}({\cal W})$ has the form: \beqa
\mathcal{W}_{\rm fcn} &=&\left[
\begin{array}{cc}
\cos^{2}\theta +|\alpha|^{2}\sin^{2}\theta & \alpha^{*}\beta
\sin^{2}\theta  \\
\alpha \beta ^{*}\sin^{2}\theta  & |\beta|^{2}\sin^{2}\theta
\end{array}
\right]. \eeqa

These results serve as background in addressing polarization of
non-paraxial or non-beam light fields, i.e., the three dimensional
fields expressed as given in (\ref{E}), and non-trivially entangled.
We recall the observation  that (\ref{E}) has the character of an
entangled description of two parties jointly, where the ``parties"
here are the two distinct aspects (or degrees of freedom) of the
light field, namely the lab space direction of the optical field and
the statistical function space characterizing the strength of the
field. This suggests making a Schmidt-type analysis
\cite{SchmidtThm, Eberly06}, as frequently used in discussions of
quantum entanglement data in few-mode and multi-mode photonic
contexts \cite{Law-etal00, Law-Eberly04, Boyd-Howell05, Woerdman05,
Fedorov-etal05}.

The Schmidt theorem applies to any kind of two-party vector, whether
quantum or not, and we begin by calculating the same two reduced
matrices ${\cal W}_{\rm lab}$ and ${\cal W}_{\rm fcn}$ defined
above, except that now they are three-dimensional. Since they arise
from a common hermitean ${\cal W}$, which is the $3 \times 3$ analog
of (\ref{W-e}), they share the same three (real) eigenvalues,
$\ka1^2, \ka2^2, \ka3^2$. Their eigenvectors \cite{evaleqns} are
orthonormal and occur in pairs. The Schmidt-form result enables the
optical field to be written immediately in a perfectly organized
form in terms of these eigenvalues and eigenvectors: \beq
\label{Eoptimum} |{\bE}\ra/\sqrt{I} = \ka1 |{\bu}_1\ra|f_1\ra + \ka2
|{\bu}_2\ra|f_2\ra +\ka3 |{\bu}_3\ra|f_3\ra. \eeq Here perfect
organization means that orthogonality conditions apply in both
vector spaces at the same time: $\la f_i|f_j\ra = \la\bu_i|\bu_j\ra
= \delta_{ij}$, and since intensity has been factored out, the three
$\kappa$s are normalized on the surface of a unit sphere: $\ka1^2 +
\ka2^2 + \ka3^2 = 1$.

This Schmidt decomposition of the field is unique up to a rotation
at most, and allows interesting questions to be answered by
inspection. For example, it is obvious that (\ref{Eoptimum}) can
take the  completely polarized (fully factored and so non-entangled)
form $\bE = \bu F = \sqrt{I}\,\bu\,f$ only when two of the $\kappa$s
have the value zero. Because the three  $|f\ra$ vectors are mutually
statistical orthogonal, no other $\kappa$ values can produce a
completely polarized result. For such a field, no matter which lab
direction ${\bf v}$ is used for a projective measurement of ${\bf
E}$, the statistical-functional features of that component will
always be exactly those of the function $f$. And similarly, no
matter what projection in function space is employed, the projected
field's direction will always be ${\bf u}$ in lab space.

The connection to standard polarization measures is not complicated.
Obviously the conventional definition \cite{statoptics} of degree of
polarization $P$, designed for a beam-type field, is sensible only
when there is zero projection along one of the three $u$'s, say
along $u_3$, in which case $\ka3 = 0$. Then $P$ has several compact
expressions \cite{statoptics, Brosseau, Fedorov11}: \beq
\label{2dimP} P^2 = 1- 4Det({\cal W}_{\rm lab}) =
|\kappa_1^2-\kappa_2^2|^2= 1-2(1-1/K) , \eeq where $K$ will be
introduced below.  As is well known \cite{Brosseau}, the beam-based
definition resists being generalized to the three dimensional case
when none of the $\kappa_i$s is zero in (\ref{Eoptimum}). In
addition, with $\ka3 = 0$, the field cannot point even slightly into
the direction ${\bu}_3$. Thus to call it ``unpolarized" is not fully
sensible even if $\ka1 = \ka2$, making $P$ zero.

The Schmidt decomposition automatically provides a very useful
``weight" parameter $K$ \cite{Grobe-etal}, which counts the
non-integer effective number of dimensions  needed by the optical
field. The expression for $K$ is \beq \label{K} K = 1/[{\kappa_1^{4}
+\kappa_{2}^{4} + \kappa_{3}^{4} }], \eeq which gives greater weight
to the vector directions with the larger absolute $\kappa$ values.
It's easy to see that $K$ lies between 1 and 3 and incorporates the
beam-type field case automatically when any one of the three
$\kappa$'s is zero.

\begin{figure}[t!]
\includegraphics[width= 5.5 cm]{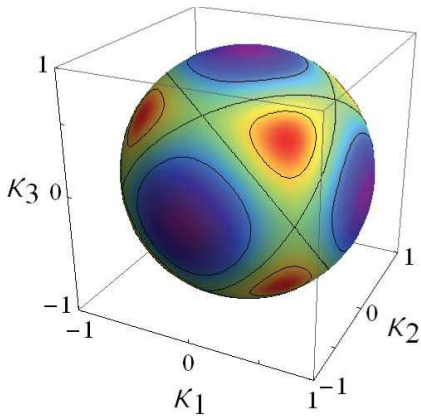}
\caption{The entanglement measure $K$  varies from 1 to 3 over the
unit polarization sphere.  The purple zone-centers touch the
surfaces of the cube ($K$=1, completely polarized), and the red
centers are completely unpolarized ($K$=3).  Planar-polarized fields
($K=2$ or $P$=0) are located at the corners of the triangular
regions where one $\kappa_j^2=0$ and the others equal $1/{2}$. The
mesh lines locate partially polarized values $K = 3/2, 2, 5/2$.}
\label{f.polarsphere}
\end{figure}

$K=1$ occurs if $\ka2 = \ka3 = 0$, which signals a one-term ${\bf
E}$, i.e., completely polarized light, obtainable only if the
original field components could have been rotated into a fully
factored form: ${\bf E} = {\bf u}F$. This is of course generally not
possible. The farthest departure occurs for the value $K=3$, when
$\kappa_j^2 = 1/3$ for all three components, and all $f_j$ are
equally intense. This is maximal entanglement and also what is
sensibly called a completely unpolarized field. Intermediate values
of $K$ represent intermediate degrees of entanglement (partially
polarized light). All of these conditions are associated with
$\kappa$ values that identify points on the unit sphere in Fig.
\ref{f.polarsphere}.

In summary we have reformulated polarization theory as entanglement
analysis. The Schmidt theorem approach automatically provides an
optimum expression for any light field by identifying its orthogonal
directions in lab space, ${\bf u}_{i}$, and its associated
amplitudes $f_{i}$ in statistical function space. This is possible
because every optical field is a quantity existing simultaneously in
those two independent vector spaces. The interpretation of degree of
polarization naturally corresponds to the separability between the
two spaces for both planar and non-planar cases. In this new
perspective, polarization is a characterization of the correlation
between the vector nature and the statistical nature of the light
field.

We acknowledge the benefit of discussions with M.A. Alonso, T.G.
Brown, G. Leuchs, M. Lahiri, and E. Wolf, as well as financial
support from NSF PHY-0855701.


\begin{thebibliography}{99}
\bibitem{evaleqns} The eigenvalue equations are just $3 \times 3$ matrix
equations: ${\cal W}_{\rm lab}|{\bu}_j\ra = \kappa_j^2|{\bu}_j\ra$
and ${\cal W}_{\rm fcn}|{f}_j\ra = \kappa_j^2|{f}_j\ra$.
\end{thebibliography}
\end{document}